\documentclass[11pt]{article}
\usepackage{acl2015}
\usepackage{times}
\usepackage{url}
\usepackage{latexsym}
\usepackage{graphicx}
\usepackage{multicol}

\title{Leverage Financial News to Predict Stock Price Movements \\ Using Word Embeddings and Deep Neural Networks}

\author{Yangtuo Peng \and Hui Jiang \\ 
Department of Electrical Engineering and Computer Science \\
York University, 4700 Keele Street, Toronto, Ontario, M3J 1P3, Canada \\
\em emails: tim@cse.yorku.ca, hj@cse.yorku.ca, 
}

\date{}

\begin{document}
\maketitle
\begin{abstract}
Financial news contains useful information on public companies and the market. In this paper we apply the popular word embedding methods and deep neural networks to leverage financial news to predict stock price movements in the market. Experimental results have shown that our proposed methods are simple but very effective, which can significantly improve the stock prediction accuracy on a standard financial database over the baseline system using only the historical price information. 
\end{abstract}

\section{Introduction}

In the past few years, deep neural networks (DNNs) have achieved huge successes in many data modeling and prediction tasks, ranging from speech recognition, computer vision to natural language processing. In this paper, we are interested in applying the powerful deep learning methods to financial data modeling to predict stock price movements. 

Traditionally neural networks have been used to model stock prices as time series for the forecasting purpose, such as in \cite{Kaastra1996,Adya1998,Chan2000,Skabar2002,Zhu2008}. In these earlier work, due to the limited  training data and computing power available back then, normally shallow neural networks were used to model various types of features extracted from stock price data sets, such as historical prices, trading volumes, etc, in order to predict future stock yields and market returns. 
More recently, in the community of natural language processing, many methods have been proposed to explore  additional information (mainly online text data) for stock forecasting, such as financial news \cite{xie-EtAl:2013:ACL2013,ding-EtAl:2014:EMNLP2014}, twitters sentiments \cite{si-EtAl:2013:Short,si-EtAl:2014:EMNLP2014}, microblogs \cite{barhaim-EtAl:2011:EMNLP}. 
For example, \cite{xie-EtAl:2013:ACL2013} propose to use semantic frame parsers to generalize from sentences to scenarios to detect the (positive or negative) roles of specific companies, where support vector machines with tree kernels are used as predictive models. On the other hand, \cite{ding-EtAl:2014:EMNLP2014} propose to use various lexical and syntactic constraints to extract event features for stock forecasting, where they have investigate both linear classifiers and deep neural networks as predictive models. 

In this paper, we propose to use the recent word embedding methods \cite{mikolov2013distributed} to select features from on-line financial news corpora, and employ deep neural networks (DNNs) to predict the future stock movements based on the extracted features. Experimental results have shown that the features derived from financial news are very useful and they can significantly improve the prediction accuracy over the baseline system that only relies on the historical price information. 


\section{Our Approach}

In this paper, we use deep neural networks (DNNs) as our predictive model, which takes as input the features extracted from both historical price information and on-line financial news to predict the stock movements in the future (either up or down).
%

\subsection{Deep Neural Networks}

The structure of DNNs used in this paper is a conventional multi-layer perceptron with many hidden layers. An $L$-layer DNN consisting of $L-1$ hidden nonlinear layers and one output layer. The output layer is used to model the posterior probability of each output target.
In this paper, we use the rectified linear activation function, i.e., $f(x)=\max(0, x)$, to compute from activations to outputs in each hidden layer, which are in turn fed to the next layer as inputs.
For the output layer, we use the {\em softmax} function to compute posterior probabilities between two nodes, standing for {\em stock-up} and {\em stock-down}.


\subsection{Features from historical price data}
\label{sec_historical}

In this paper, for each target stock on a target date, we choose the previous five days' closing prices and concatenate them to form an input feature vector for DNNs: $P = (p_{t-5}, p_{t-4}, p_{t-3}, p_{t-2}, p_{t-1})$, where $t$ denotes the target date, and $p_{m}$ denotes the closing price on the date $m$. We then normalize all prices by the mean and variance calculated from all closing prices of this stock in the training set. 
In addition, we also compute first and second order differences among the five days' closing prices, which are appended as extra feature vectors. For example, we compute the first order difference as follows:
$\Delta P =  (p_{t-4}, p_{t-3}, p_{t-2}, p_{t-1})$ $ - (p_{t-5}, p_{t-4}, p_{t-3}, p_{n-2})$. In the same way, the second order difference is calculated by taking the difference between two adjacent values in each $\Delta P$. Finally, for each target stock on a particular date, the feature vector representing the historical price information consists of 
$P$, $\Delta P$ and $\Delta\Delta P$.

\subsection{Financial news features}
\label{sec_financial_news}

In order to extract fixed-size features suitable to DNNs from financial news corpora, we need to pre-process the text data. For all financial articles, we first split them into sentences. We only keep those sentences that mention at least one stock name or one public company. Each sentence is labelled by the publication date of the original article and the mentioned stock name. It is possible that multiple stocks are mentioned in one sentence. In this case, this sentence is labeled several times for each mentioned stock. 
We then group these sentences by the  publication dates and the underlying stock names to form the samples. 
Each sample contains a list of sentences that were published on the same date and mentioned the same  stock or company. Moreover, each sample is labelled as {\em positive} (``price-up'') or {\em negative} (``price-down'') based on its next day's closing price consulted from the CRSP financial database \cite{CRSP-book}.
In the following, we introduce our method to extract three types of features from each sample. 

{\bf (1) Bag of keywords (BoK)}: 
We first select the keywords based on the recent word embedding methods in \cite{mikolov2013efficient,mikolov2013distributed}. Using the popular word2vec method from Google\footnote{https://code.google.com/p/word2vec/}, we first 
compute the vector representations for all words occurring in the training set.
Secondly, we manually select a small set of seed words, i.e., nine words of \{{\em surge}, {\em rise}, {\em shrink}, {\em jump}, {\em drop}, {\em fall}, {\em plunge}, {\em gain}, {\em slump}\} in this work, which are believed to have a strong indication to the stock price movements. 
Next, these seed words are used to search for other useful keywords based on the cosine distances calculated between the word vector of each seed word and that of other words occurring in the training set. For example, based on the pre-calculated word vectors, we have found other words, such as {\em rebound}, {\em decline}, {\em tumble}, {\em slowdown}, {\em climb}, which are very close to at least one of the seed words. In this way, we have searched all words occurring in training set and kept the top 1,000 words (including the nine seed words) as the keywords for our prediction task. Finally, a 1000-dimension feature vector, called {\em bag-of-keywords} or {\em BoK}, is generated for each sample. Each dimension of the {\em BoK} vector is the {\em TFIDF} score computed for each selected keyword from the whole training corpus. 


{\bf (2) Polarity score (PS)}: 
We further compute so-called {\em polarity} scores \cite{Turney:2003:MPC:944012.944013,Turney:2010:FMV:1861751.1861756} to measure how each keyword is related to stock movements and how each keyword applies to a target stock in each sentence.  To do this, we first compute the point-wise mutual information for each keyword $w$: 
$
\mbox{PMI}(w, pos) = \log \frac{\mbox{freq}(w, pos) \times N}{\mbox{freq}(w) \times \mbox{freq}(pos)},
$
where $\mbox{freq}(w, pos)$ denotes the frequency of the keyword $w$ occurring in all positive samples, $N$ denotes the total number of samples in the training set, $\mbox{freq}(w)$ denotes the total number of keyword $w$ occurring in the whole training set and $\mbox{freq}(pos)$ denotes the total number of positive samples in the training set. Furthermore, we calculate the polarity score for each keyword $w$ as:
$
\mbox{PS}(w) = \mbox{PMI}(w, pos) - \mbox{PMI}(w, neg).
$
Obviously, the above polarity score $\mbox{PS}(w)$ measures how (either positively or negatively) each keyword is related to stock movements and by how much. 

Next, for each sentence in all samples, we need to detect how each keyword is related to the mentioned stock. To do this,
we use the Stanford parser \cite{StanfordParser2006} to detect whether the target stock is a subject of the keyword or not. If the target stock is not the subject of the keyword in the sentence, we assume the keyword is {\em oppositely } related to the underlying stock. As a result, we need to flip the sign of the polarity score. Otherwise, if the target stock is the subject of the keyword, we keep the keyword's polarity score as it is. For example, in a sentence like {\em ``Apple slipped behind Samsung and Microsoft in a 2013 customer experience survey from Forrester Research''}, we first identify the keyword {\em slipped}, based on the parsing result, we know {\em Apple} is the subject while  {\em Samsung} and {\em Microsoft} are not. Therefore, if this sentence is used as a sample for {\em Apple}, the above polarity score of ``{\em slipped}'' is directly used. However, if this sentence is used as a sample for {\em Samsung} or {\em Microsoft}, the polarity score of ``{\em slipped}'' is flipped by multiplying $-1$.

Finally, the resultant polarity scores are multiplied to the {\em TFIDF} scores to generate another 1000-dimension feature vector for each sample.
%

{\bf (3) Category tag (CT)}: 
We further define a list of categories that may indicate a specific event or activity of a public company, which we call as category tags. In this paper, the defined category tags include: {\em new-product}, {\em acquisition}, {\em price-rise}, {\em price-drop}, {\em law-suit}, {\em fiscal-report}, {\em investment}, {\em bankrupt}, {\em government}, {\em analyst-highlights}. 
Each category is first manually assigned with a few words that are closely related to the category. 
For example, we have chosen {\em released}, {\em publish}, {\em presented}, {\em unveil} as a list of seed words for the category  {\em new-product},  which indicates the company's announcement of new products. 
Similarly, we use the above word embedding model to automatically expand the above word list by searching for more words that have closer cosine distances with the selected seed words. In this paper, we choose the top 100 words to assign to each category. 

After we have collected all key words for each category, for each sample, we count the total number of occurrences of all words under each category, and then we take the logarithm to obtain a feature vector as
${V} = (\log N_{1}, \log N_{2}, \log N_{3}, ..., \log N_{c})$,
where 
$N_{c}$ denotes the total number of times the words in category $c$ appear in a sample. 

\subsection{Predicting Unseen Stocks via Correlation Graph}
\label{section-correlation}

\begin{figure}[t]
 	\centering
 	\includegraphics[width=0.85\linewidth,height=0.55\linewidth]{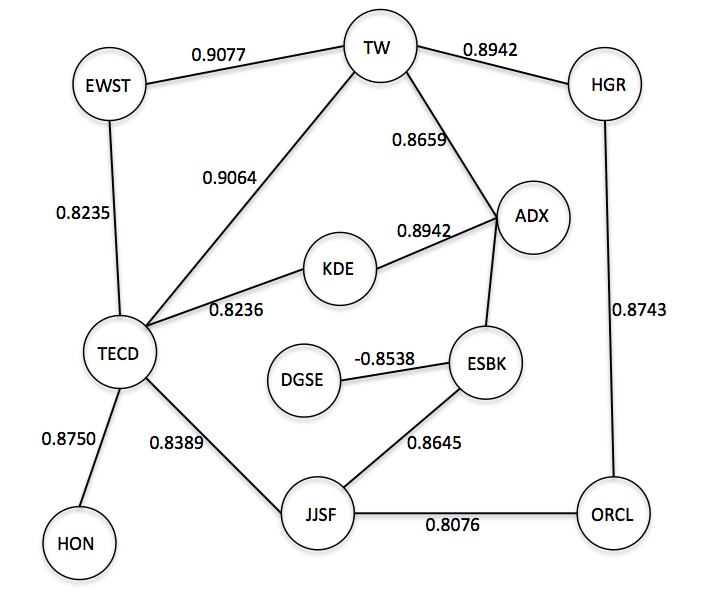}
 	\caption{Illustration of a part of correlation graph \label{figure-correlation-graph}}
\end{figure}

There are a large number of stocks trading in the market. However, we normally can only find a fraction of them mentioned in daily financial news. Hence, for each date, the above method can only predict those stocks mentioned in the news. In this section, we propose a new method to extend to predict more stocks that may not be directly mentioned in the financial news. 
Here we propose to use a stock correlation graph, shown in Figure \ref{figure-correlation-graph}, to predict those unseen stocks. The stock correlation graph is an undirected graph, where each node represents a stock and the arc between two nodes represents the correlation between these two stocks.  For example, if some stocks in the graph are mentioned in the news on a particular day, we first use the above method to predict these mentioned stocks. Afterwards, the predictions are propagated along the arcs in the graph to generate predictions for those unseen stocks. 
 
(1) Build the graph: We choose the top 5,000 stocks from the CRSP database \cite{CRSP-book} to construct the correlation graph. At each time, any two stocks in the collection are selected to align their closing prices based on the related dates (between 2006/01/01 - 2012/12/31). Then we calculate the correlation coefficient between  the closing prices of these two stocks.
The computed correlation coefficient (between $-1$ and $1$) is attached to the arc connecting these two stocks in the graph, indicating their price correlation. The correlation coefficients are calculated for every pair of stocks from the collection of 5,000 stocks. In this paper we only keep the arcs with an absolute correlation value greater than $0.8$, all other edges are considered to be unreliable and pruned from the graph, a tiny fraction of which is shown in Figure  \ref{figure-correlation-graph}.

(2) Predict unseen stocks: In order to predict price movements of unseen stocks, we first take the prediction results of those mentioned stocks from the DNN outputs, by which we construct a 5000-dimension vector ${\bf x}$. Each dimension of ${\bf x}$ corresponds to one stock and we set zeros for all unseen stocks. The above graph propagation process can be mathematically represented as a matrix multiplication: 
${\bf x}' =  {\bf A} {\bf x}$,
where ${\bf A}$ is a symmetric matrix denoting all correlation weights in the graph. Of course, the graph propagation, i.e. matrix multiplication, may be repeated for several times until the prediction ${\bf x}'$ converges. 


\section{Dataset}

The financial news data we used in this paper are provided by \cite{ding-EtAl:2014:EMNLP2014} which contains 106,521 articles from Reuters and 447,145 from Bloomberg. The news articles were published in the time period from October 2006 to December 2013. The historical stock security data are obtained from the Centre for Research in Security Prices (CRSP) database \cite{CRSP-book}. We only use the security data from 2006 to 2013 to match the time period of the financial news.  Base on the samples' publication dates, we split the dataset into three sets: a training set (all samples between 2006-10-01 and 2012-12-31), a validation set (2013-01-01 and 2013-06-15) and a test set (2013-06-16 to 2013-12-31). The training set containts 65,646 samples, the validation set 10,941 samples, and the test set 9,911 samples.

\section{Experiments}

\subsection{Stock Prediction using DNNs}

In the first set of experiments, we use DNNs to predict stock's price movement based on a variety of features, namely producing a polar prediction of the price movement on next day (either {\em price-up} or {\em price-down}).
Here we have trained a set of DNNs using different combinations of feature vectors and found that the DNN structure of 4 hidden layers (with 1024 hidden nodes in each layer) yields the best performance in the validation set. We use the historical price feature alone to create the baseline and various features derived from the financial news are added on top of it. 
We measure the final performance by calculating the error rate on the test set.  
As shown in Table \ref{table-stock-prediction}, the features derived from financial news can significantly improve the prediction accuracy and we have obtained the best performance (an error rate of 43.13\%) by using all the features discussed in Sections \ref{sec_historical} and \ref{sec_financial_news}.

\begin{table}[h]
\begin{center}
\begin{tabular}{|l|l|}
\hline \em feature combination & \em error rate \\
\hline price & 48.12\% \\
\hline price + BoK & 46.02\% \\
\hline price + BoK + PS & 43.96\% \\
\hline price + BOK + CT & 45.86\% \\
\hline price + PS & 45.00\% \\
\hline price + CT & 46.10\% \\
\hline price + PS +CT & 46.03\% \\
\hline price + BoK + PS + CT& {\bf 43.13\%} \\
\hline
\end{tabular}
\end{center}
\caption{\label{table-stock-prediction} Stock prediction error rates on the test set.}
\end{table}

\subsection{Predict Unseen Stocks via Correlation}

\begin{figure}[t]
	\centering
	\includegraphics[width=0.85\linewidth,height=0.55\linewidth]{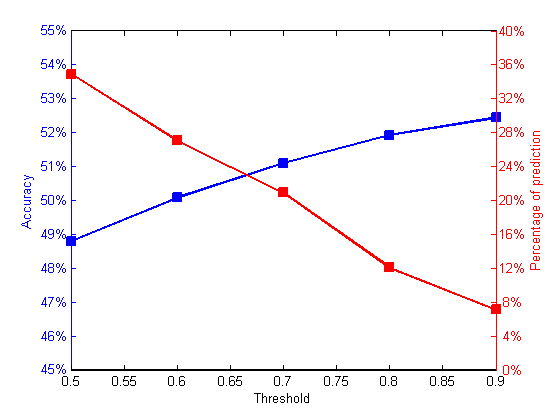}
	\caption{Predict unseen stocks via correlation}\label{figure-unseen-stocks}
\end{figure}

Here we group all outputs from DNNs based on the dates of all samples on the test set. 
For each date, we create a vector ${\bf x}$ based on the DNN prediction results for all observed stocks and zeros for all unseen stocks, as described in section \ref{section-correlation}. Then, the vector is propagated through the correlation graph to 
generate another set of stock movement prediction. We may apply a threshold on the propagated vector to prune all low-confidence predictions. The remaining ones may be used to predict some stocks unseen on the test set. The prediction of all unseen stocks is compared with the actual stock movement on next day. 
Experimental results are shown in Figure \ref{figure-unseen-stocks}, where the left y-axis denotes the prediction accuracy and the right y-axis denotes the percentage of stocks predicated out of all 5000 per day under each pruning threshold. For example, using a large threshold (0.9), we may predict with an accuracy of 52.44\% on 354 extra unseen stocks per day, in addition to predicting only 110 stocks per day on the test set. 

 
\section{Conclusion}

In this paper, we have proposed a simple method to leverage financial news to predict stock movements based on the popular word embedding and deep learning techniques. Our experiments have shown that the financial news is very useful in stock prediction and the proposed methods can significantly improve the prediction accuracy on a standard financial data set.

\section*{Acknowledgments}

This work was supported in part by an NSERC Engage grant from Canadian federal government.

\bibliographystyle{acl}
\bibliography{emnlp2015_yangtuo}









\end{document}